\documentclass{PoS}

\title{Quark mass dependence of quarkonium properties at finite temperature}

\ShortTitle{Quark mass dependence of quarkonium properties at finite temperature}

\author{\speaker{H.~Ohno}$^{ab}$, H.-T.~Ding$^c$, and O.~Kaczmarek$^d$\\
        \llap{$^a$}Center for Computational Sciences, University of Tsukuba, Tsukuba, Ibaraki 305-8577, Japan\\
        E-mail: \email{hohno@ccs.tsukuba.ac.jp} \\
        \llap{$^b$}Physics Department, Brookhaven National Laboratory, Upton, NY 11973, USA \\
        \llap{$^c$}Key Laboratory of Quark \& Lepton Physics (MOE) and Institute of Particle Physics,\\
        Central China Normal University, Wuhan, 430079, China\\
        E-mail: \email{hengtong.ding@mail.ccnu.edu.cn} \\
        \llap{$^d$}Fakult\"{a}t f\"{u}r Physik, Universit\"{a}t Bielefeld, D-33501 Bielefeld, Germany \\
        E-mail: \email{okacz@physik.uni-bielefeld.de}}

\abstract{Quarkonium properties at finite temperature have been studied with quark masses of the charm and bottom quarks.
Our simulations have been performed in quenched QCD with the $O(a)$-improved Wilson quarks on large and fine isotropic lattices
with the spatial lattice extents $N_\sigma=$ 96, 192 and the corresponding lattice spacings $a=$ 0.0190, 0.00967 fm, respectively,
at temperatures in a range between about 0.7$T_c$ and 1.4$T_c$. We show temperature and quark mass dependence of quarkonium correlation
functions and related physical quantities: the quark number susceptibility and the heavy quark diffusion constant.}

\FullConference{The 32nd International Symposium on Lattice Field Theory\\
                 23-28 June, 2014\\
                 Columbia University New York, NY}

\begin{document}

\section{Introduction}

The behavior of quarkonia in a hot medium is of great interest to understand properties of the quark-gluon plasma (QGP) expected
to form in relativistic heavy ion collisions. Because of the color Debye screening in the deconfined phase
quarkonia should melt at certain dissociation temperatures. Therefore suppression of quarkonium yields can be
an important signal of QGP formation in experiments \cite{Matsui:1986dk}. Actually observation of the J/$\psi$ suppression
at SPS \cite{Arnaldi:2009ph}, RHIC \cite{Adare:2008qa} and LHC \cite{Abelev:2012rv,Aad:2010aa,Chatrchyan:2012np}
as well as the sequential $\Upsilon$ suppression at LHC \cite{Chatrchyan:2012lxa} have been reported. However, since there
are cold nuclear matter effects which can also contribute to the quarkonium suppression, it is important to have good theoretical
understanding of in-medium properties of quarkonia.

Not only in-medium behavior of bound states but also that of single heavy quark is interesting
since the elliptic flow of heavy quarks has been observed in the experiment \cite{Adare:2006nq}.
This gives an evidence of collective motion due to hydrodynamic effects. Here the heavy quark diffusion constant
related to energy loss of an in-medium heavy quark and the ratio of shear viscosity to entropy density
is one of important quantities to explain this hydrodynamic aspect of the experimental results.
Although there are several phenomenological studies about heavy quark diffusion \cite{Moore:2004tg,CaronHuot:2007gq},
their results are not yet conclusive. Therefore calculation based on the first principle is needed.

Lattice QCD is a powerful approach to study physics where non-perturbative effects in the strong interaction are important.
Thus many lattice QCD studies on charmonia at finite temperature have already been done \cite{Jakovac:2006sf,Aarts:2007pk,Ohno:2011zc,Ding:2012sp},
however, their results on dissociation temperatures are still controversial.
Recently studies on bottomonia have also been reported in the framework of NRQCD \cite{Aarts:2010ek,Aarts:2012ka}.
The heavy quark diffusion constant has been estimated in \cite{Ding:2012sp} at one lattice spacing.
We have reported our recent study on continuum estimate of the heavy quark momentum diffusion in context of a heavy quark effective theory \cite{Kaczmarek:2014jga}.
In this contribution we show correlation functions of both charmonia and bottomonia computed in quenched lattice QCD at finite
temperature and discuss the behavior of charmonia and bottomonia in a hot medium.
We mainly forcus on the changes of the correlation functions at temperatutres from 0.7$T_c$ to 1.4$T_c$.
Results at 1.1$T_c$ and 1.2$T_c$ have been reported in our previous study\cite{Ohno:2013rka}.
We also show the quark number susceptibility and heavy quark diffusion constant which can be calculated from correlation functions in the vector channel.
One of our goals is to calculate the quantities mentioned above in the continuum limit.
However we only show results given at two different lattice spacings in this contribution so far.

\section{Quarkonium correlation and spectral functions}

In this study we investigate the Euclidean meson correlation function at vanishing momentum, which is defined as
\begin{equation}
G_H(\tau) \equiv \int \langle J_H(\vec{x},\tau)J_H(\vec{0},0) \rangle d^3x,
\end{equation}
where $J_H(\vec{x},\tau) \equiv \bar{q}(\vec{x},\tau)\Gamma_H q(\vec{x},\tau)$ is a meson operator and
$\Gamma_H = \gamma_5,\;\gamma_\mu,\;\textrm{\boldmath $1$},\; \gamma_\mu \gamma_5$ corresponds to pseudo-scalar (PS), vector (V),
scalar (S) and axial-vector (AV) channels, respectively. Then, at a certain temperature $T$, the spectral function $\rho_H(\omega)$,
which has all information about in-medium properties of the quarkonium, is related to the correlation function by
\begin{equation}
G_H(\tau,T) = \int^\infty_0 \frac{d\omega}{2\pi} \rho_H(\omega)\frac{\cosh(\omega(\tau-1/2T))}{\sinh(\omega/2T)}.
\end{equation}
To investigate the temperature dependence of the spectral function, instead of estimating the spectral function itself from the correlation function,
we consider the reconstructed correlation function defined by
the kernel and the spectral function at temperatures $T$ and $T^{\prime}$, respectively, as
\begin{equation}
G_\mathrm{rec}(\tau,T;T^\prime) \equiv \int^\infty_0\frac{d\omega}{2\pi} \rho(\omega,T^{\prime})\frac{\cosh(\omega(\tau-1/2T))}{\sinh(\omega/2T)}.
\end{equation}
Since $G(\tau,T)$ and $G_{\mathrm{rec}}(\tau,T;T^\prime)$ have same trivial temperature dependence coming from the kernel, differences between these two
quantities originate from thermal modifications of the spectral function. Here, since we do not calculate the spectral function itself, instead
we calculate the reconstructed correlation function by using a relation \cite{Ding:2012sp} as follows:
\begin{equation}
G_\mathrm{rec}(\tau,T;T^\prime) = \sum^{1/T^\prime-1/T+\tau}_{\tau^\prime=\tau,\;\Delta\tau^\prime=1/T} G(\tau^\prime,T^\prime),
\end{equation}
with $1/T = N_\tau a$ and $1/T^\prime = N^\prime_\tau a$, where $a$ is the lattice spacing and $N^\prime_\tau$ must be a multiple of $N_\tau$.
The spectral function for the vector channel is also related to the heavy quark diffusion constant $D$ as
\begin{equation}\label{diffusion}
D = \frac{1}{6\chi_{00}} \lim_{\omega \to 0} \sum^3_{i=1} \frac{\rho^{ii}_V(\omega)}{\omega},
\end{equation}
where $\chi_{00}$ is the quark number susceptibility defined by the (0,0) component of the vector correlation function as $\chi_{00} \equiv G^{00}_V(\tau)/T$
and $\rho^{ii}_V(\omega)$ is the $(i,i)$ component of the vector spectral function.

\section{Numerical results}

\begin{table}[tbp]
\begin{center}
\caption{Lattice setup. \label{t1}}
\begin{tabular}{cccccc}
\hline \hline
$\beta$ & $a$ [fm] & $N_\sigma$ & $N_\tau$ & $T/T_c$ & \# of confs. \\ \hline
7.192   & 0.0190   &  96        & 48       & 0.7     & 259          \\
        &          &            & 32       & 1.1     & 476          \\
        &          &            & 28       & 1.2     & 336          \\
        &          &            & 24       & 1.4     & 336          \\ \hline
7.793   & 0.00967  & 192        & 96       & 0.7     &  66          \\
        &          &            & 48       & 1.4     & 177          \\
\hline \hline
\end{tabular}
\end{center}
\end{table}

\begin{table}[tbp]
\begin{center}
\caption{$\kappa$ values and corresponding vector meson masses $m_V$.\label{t2}}
\begin{tabular}{ccc}
\hline \hline
$\beta$ & $\kappa$ & $m_V$ [GeV] \\ \hline
7.192   & 0.13194  & 3.105(3)    \\
        & 0.12257  & 9.468(3)    \\ \hline
7.793   & 0.13221  & 3.092(5)    \\
        & 0.12798  & 9.431(5)    \\
\hline \hline
\end{tabular}
\end{center}
\end{table}

We employed the standard plaquette gauge and $O(a)$-improved Wilson fermion actions in queched QCD.
As shown in Table \ref{t1}, we performed numerical simulations on two large and fine lattices
with the spatial lattice extents $N_\sigma =$ 96, 192 and bare lattice gauge couplings
$\beta = 6/g^2 =$ 7.192, 7.793, which correspond to lattice spacings $a =$ 0.0190, 0.00967 fm, respectively,
where the scale was determined from Sommer scale $r_0 =$ 0.49 fm \cite{Sommer:1993ce}. At temperatures in a range between about 0.7$T_c$ and 1.4$T_c$
we generated gauge configurations with an over-relaxed pseudo-heatbath algorithm. After taking more
than 1500 sweeps as thermalization, every 500 trajectories were measured. The numbers of measured configurations on each lattice
are also shown in Table \ref{t1}. To investigate the quark mass dependence we took two different $\kappa$ values as shown in Table \ref{t2},
where the quark masses were tuned such that corresponding vector meson masses $m_V$ became almost equal to experimental values of $J/\psi$
and $\Upsilon$ masses \cite{Beringer:1900zz}.

\begin{figure}[tbp]
 \begin{center}
  \includegraphics[width=49mm, angle=-90]{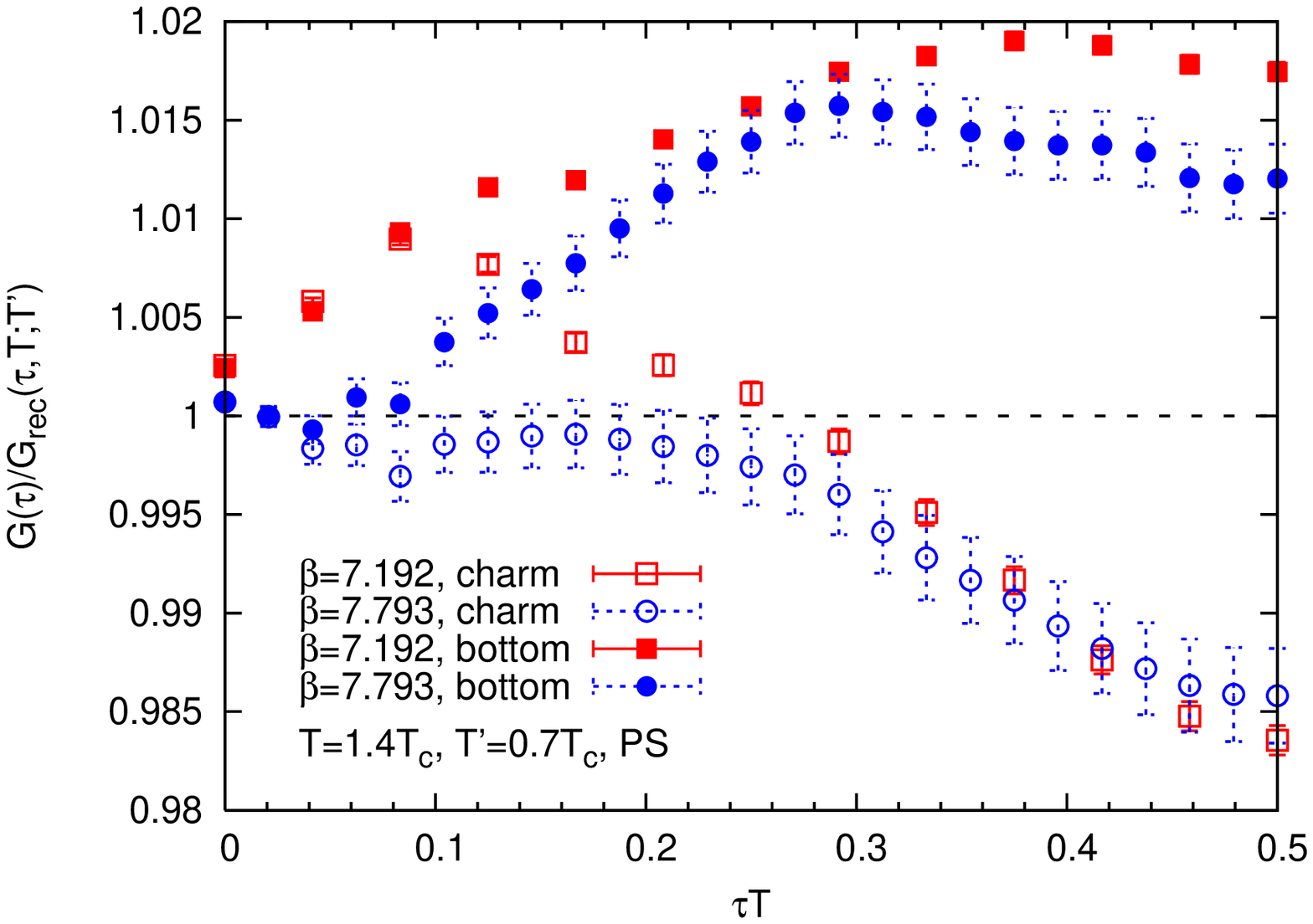}
  \includegraphics[width=49mm, angle=-90]{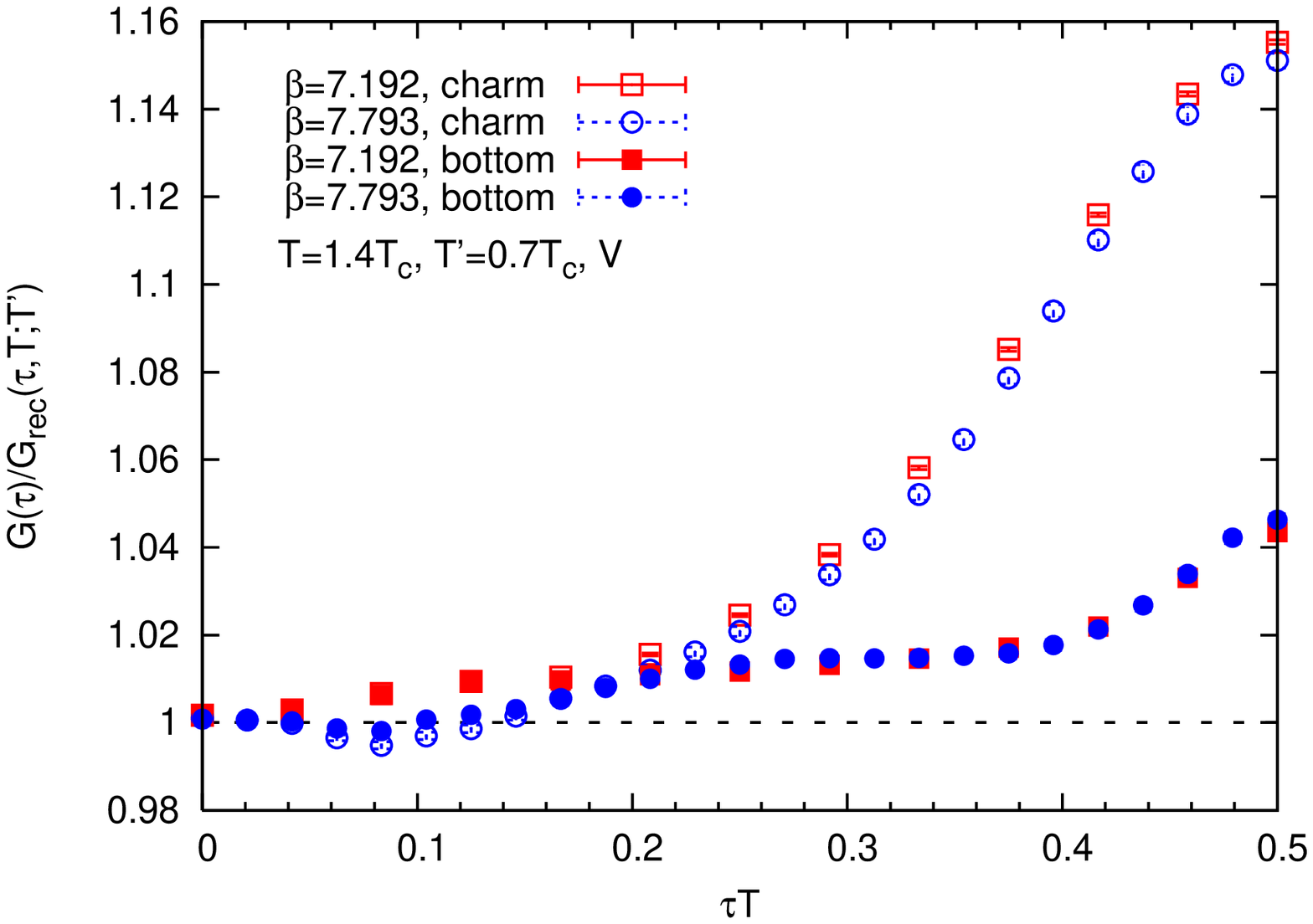}
  \includegraphics[width=49mm, angle=-90]{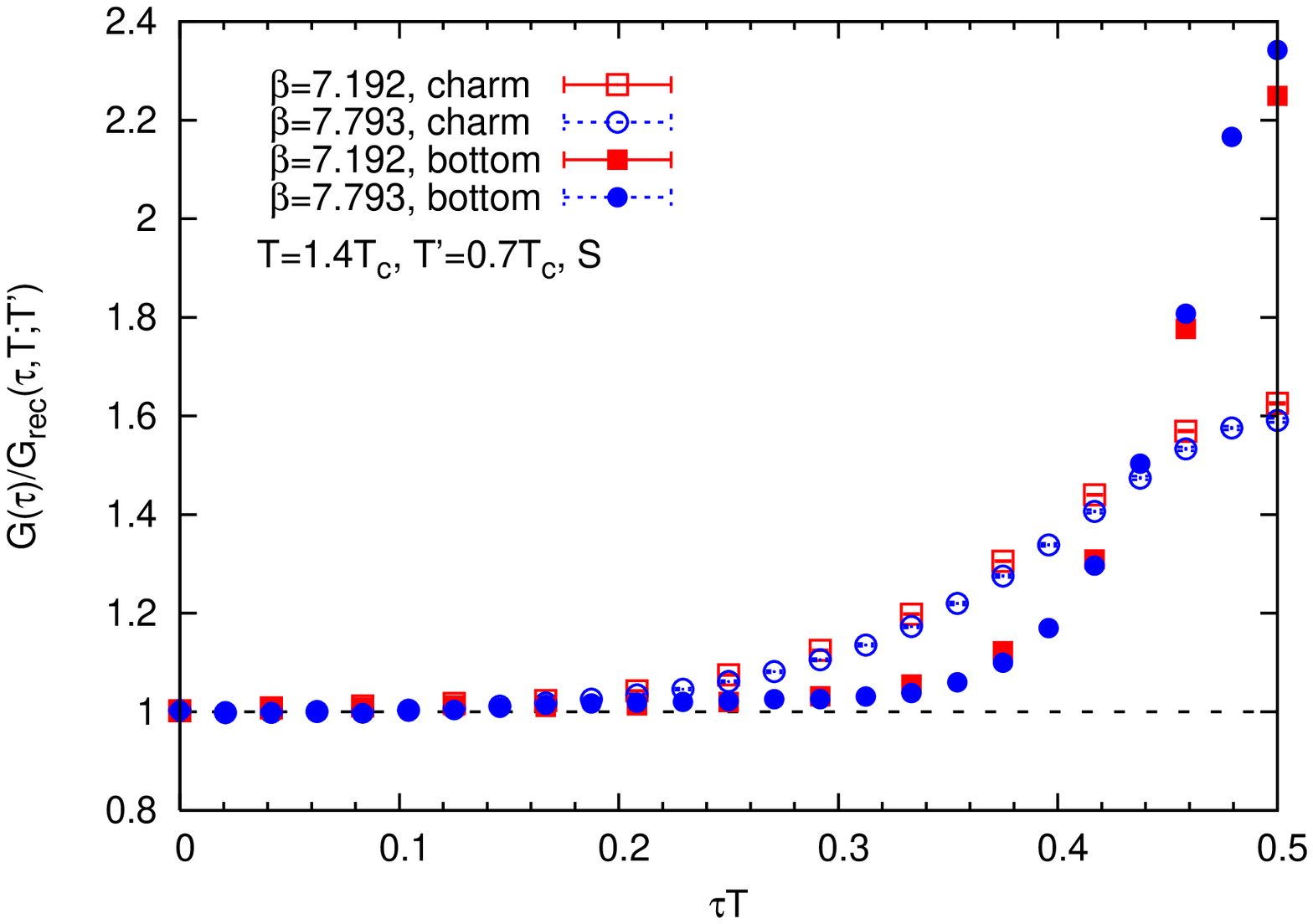}
  \includegraphics[width=49mm, angle=-90]{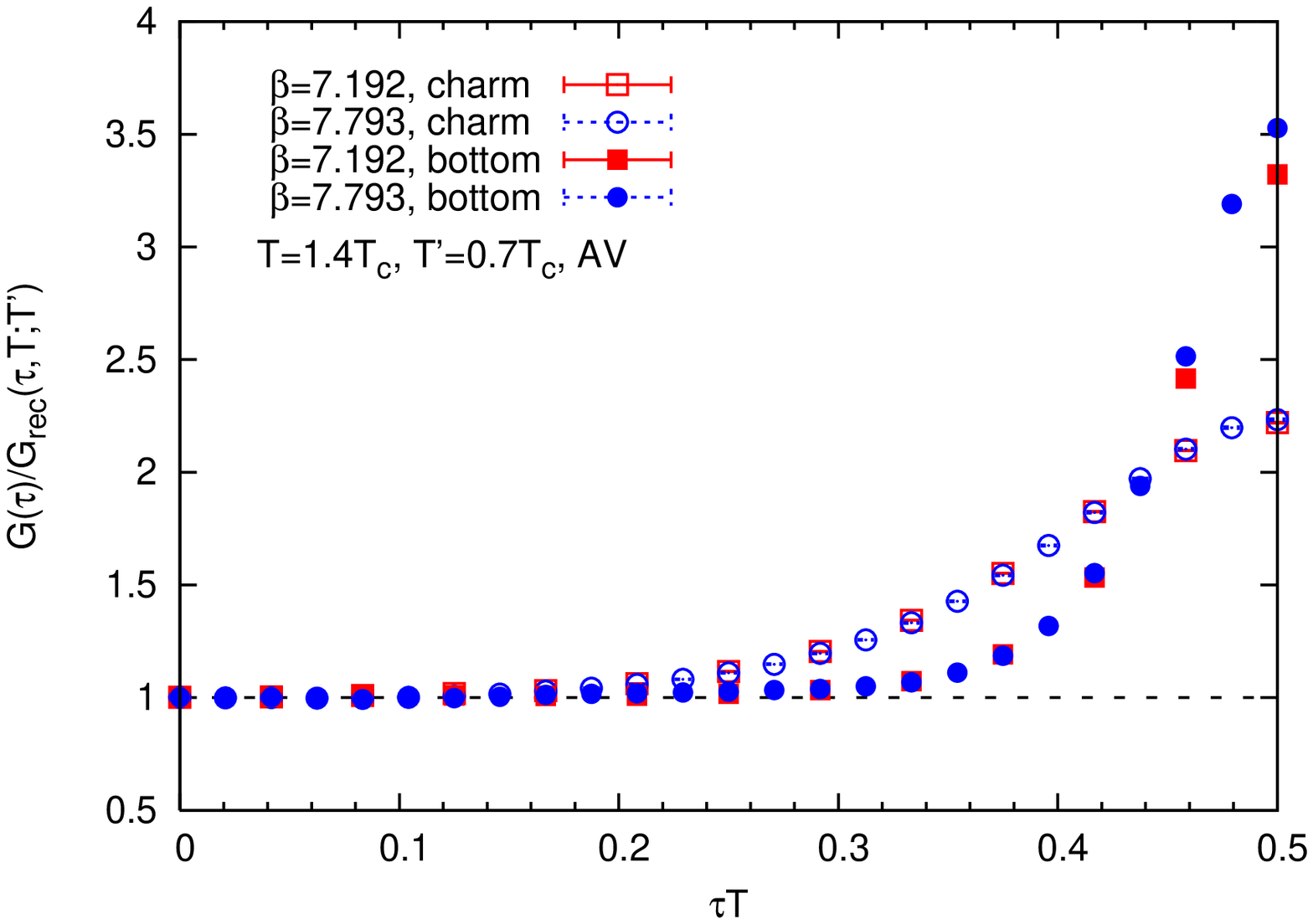}
  \caption{Ratio of ordinary to reconstructed correlation functions at $T = 1.4T_c$
  	for the PS (top-left), V (top-right), S (bottom-left) and AV (bottom-right) channels,
        where the reconstructed correlation function is calculated from the ordinary one at $T=0.7T_c$.
        Results for charmonia and bottomonia are indicated by open and filled symbols, respectively, and those at $\beta =$7.192
        and 7.793 are also shown by circle and square symbols, respectively.
        \label{rec_mcorr}}
 \end{center}
\end{figure}

In Figure \ref{rec_mcorr} ratios of the ordinary to reconstructed correlation functions at $T=1.4T_c$ are shown for the PS, V, S and AV channels,
where the reconstructed correlation function is calculated from the ordinary one at $T=0.7T_c$.
Here results for the V and AV channels are shown by averaging over the spatial polarization directions, i.e. the (1,1), (2,2) and (3,3) components.
First, it is clear to see that results for V, S and AV have strong enhancement at later time in contrast to those for the PS channel.
Looking at this enhancement in detail, it is larger for the charmonium than for the bottomonium for the V channel while it is other way
round for the S and AV channels. In our previous study \cite{Ohno:2013rka} we have found that modification of the low frequency part of the spectral
function related to the transport peak gives dominant contribution to this enhancement by seeing the midpoint-subtracted correlation
function \cite{Umeda:2007hy}. In the case of the PS channel the charmonium and bottomonium correlation functions also have quite different behavior
especially at later time, which suggests different temperature dependence of the low frequency part of their spectral functions.
Next, let us consider the cutoff effect. It can be seen that the deviation from unity at earlier time for the coarser lattice data for the PS
and V channels is due to the cutoff effect. Moreover one can also find some cutoff effect even at later time for PS channel.
Cutoff dependence looks small for S and AV channels except for data at the largest $\tau$ for charmonia.

\begin{figure}[tbp]
 \begin{center}
  \includegraphics[width=49mm, angle=-90]{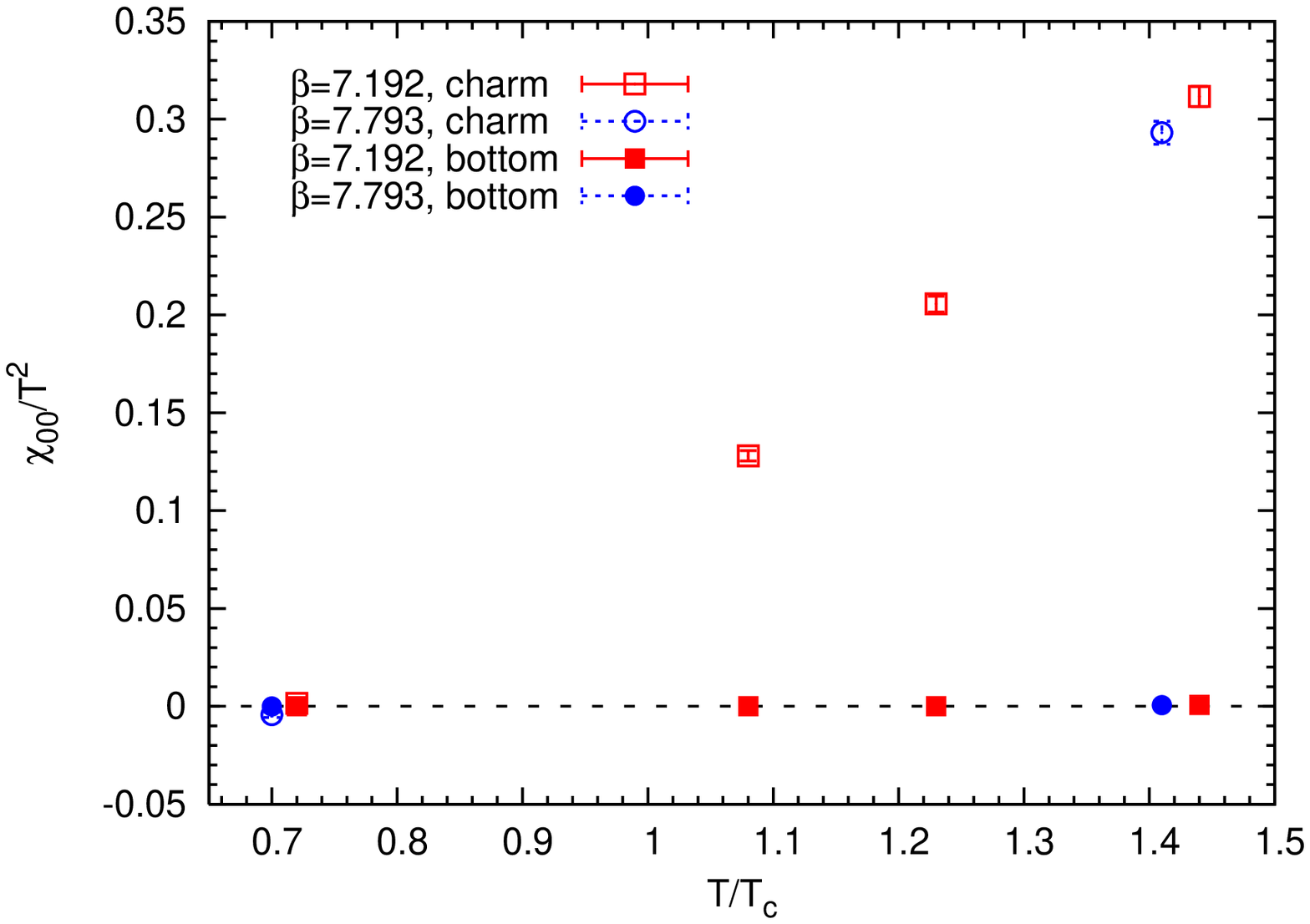}
  \includegraphics[width=49mm, angle=-90]{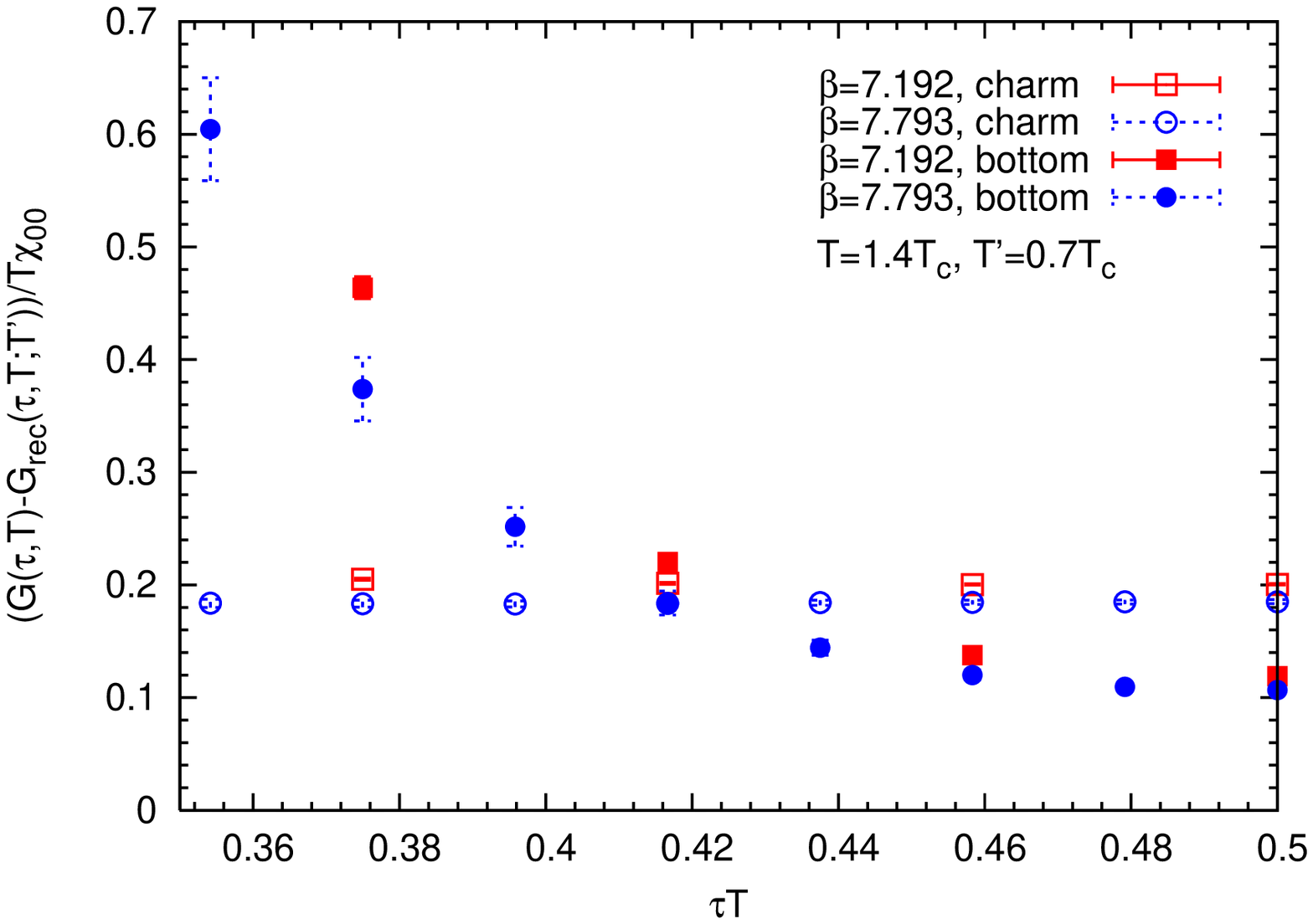}
  \caption{(Left) Temperature dependence of the quark number susceptibility normalized by temperture $\chi_{00}/T^2$.
  	(Right) Difference between the ordinary and reconstructed correlation functions normalized by the quark number susceptibility
        $\left(G(\tau,T)-G_{\mathrm{rec}}(\tau,T;T^\prime)\right)/T\chi_{00}$ at $T=1.4T_c$,
        where the reconstructed correlation function is calculated from the ordinary one at $T=0.7T_c$.
        Results for charmonia and bottomonia are indicated by open and filled symbols, respectively, and those at $\beta =$7.192
        and 7.793 are also shown by circle and square symbols, respectively.
	\label{qns_GGrecG00}}
 \end{center}
\end{figure}

On the left hand side of Figure \ref{qns_GGrecG00} the quark number susceptibility normalized by temperature $\chi_{00}/T^2$ is shown,
where $\chi_{00}$ is given by the $(0,0)$ component of the vector correlation function at $\tau T=1/2$
since it should be independent of $\tau$. It shows that cutoff effects are small and $\chi_{00}/T^2$ for the charm sector increases monotonically
as temperature increases while that for the bottom sector is almost zero at any temperature, which would suggest some thermal modification of
the charmonium states and existence of the stable bottomonium states for the V channel up to $1.4T_c$.

Here let us consider the difference between the ordinary and reconstructed correlation functions normalized by the quark number susceptibility
for the V channel, i.e. \\ $\left(G(\tau,T)-G_{\mathrm{rec}}(\tau,T;T^\prime)\right)/T\chi_{00}$, where $T=1.4T_c$ and $T^\prime=0.7T_c$.
This quantity is shown on the right hand side of Figure \ref{qns_GGrecG00}.
As we have already seen in Figure \ref{rec_mcorr}, the vector spectral function has strong thermal modification in low frequency region related
to the transport peak. If we assume that compared to the transport peak, the rest part of the spectral function has almost negligible temperature
dependence up to 1.4$T_c$, approximately $G(\tau,T)-G_{\mathrm{rec}}(\tau,T;T^\prime)$ at $\tau T=1/2$ can be written as
\begin{equation}
G(1/2T,T)-G_{\mathrm{rec}}(1/2T,T;T^\prime) \simeq \int^{\infty}_0 \frac{d\omega}{2\pi} \rho(\omega<<T)\frac{1}{\sinh(\omega/2T)},
\end{equation}
where $\rho(\omega<<T)$ consists of only the transport peak. Since the heavy quark diffusion constant $D$ is related
to the spectral function around zero frequency, according to (\ref{diffusion}), we introduce a Breit-Wigner type ansatz \cite{Petreczky:2005nh} to the transport peak as
\begin{equation}
\rho(\omega<<T) = \frac{2T\chi_{00}}{M}\frac{\omega \eta}{\omega^2 + \eta^2}, \quad \eta \equiv \frac{T}{MD},
\end{equation}
where $M\equiv ma$ is a quark mass. Thus, by solving an equation
\begin{equation}\label{diffusion2}
\frac{2}{M} \int^{\infty}_0 \frac{d\omega}{2\pi} \frac{\omega \eta}{\omega^2 + \eta^2}\frac{1}{\sinh(\omega/2T)} = \frac{G(1/2T,T)-G_{\mathrm{rec}}(1/2T,T;T^\prime)}{T\chi_{00}}
\end{equation}
with respect to $\eta$ (or $D$), one can estimate $D$ for a certain quark mass $m$.
By using $m =$ 1--2 GeV for the charm quark mass, we got $2\pi DT \simeq$ 0.6--4 and 0.5--2 at $\beta =$ 7.192 and 7.793, respectively,
which is consistent with a result given in \cite{Ding:2012sp}, although our results still have large systematic uncertainties coming from
the fit analysis and choice of quark mass.
On the other hand, we couldn't find any solution of (\ref{diffusion2}) with $m =$ 4--5 GeV in the bottom sector,
which suggests that some of our assumptions might not fit to the bottomonium case.

\section{Conclusions}

We studied charmonium and bottomonium correlation functions in quenched lattice QCD on large and fine lattices
at temperatures in a range between about 0.7$T_c$ and 1.4$T_c$ to understand in-medium properties of charmonia and bottomonia.
We investigated ratios of the ordinary to reconstructed correlation functions at 1.4$T_c$ and found large enhancement at later time
for both charmonia and bottomonia for the V, S and AV channels, which might be related to a transport peak in the spectral function.
On the other hand, the correlation functions for the PS channel didn't have such enhancement
but the behaviour in the charm and bottom sectors is different from each other, which indicates different temperature dependence
of the chamonium and bottomonium spectral functions. The cutoff dependence of the correlation functions was also checked with two different
lattice spacings and we found some cutoff effect for the PS and V channels while
that for S and AV channels looks small except for the midpoint data for charmonia.
We also investigated the quark number susceptibility, which is given from the (0,0) component of the vector correlation function.
We found that the cutoff dependence is small and the quark number susceptibilities in the charm sector increases monotonically
as temperature increases while that for the bottom sector shows almost no temperature dependence, which would suggest some thermal modification
of the charmonium states and existence of the stable bottomonium states for the V channel up to $1.4T_c$.
Finally, we considered the difference between the ordinary and reconstructed correlation functions at 1.4$T_c$, which might have dominant contributions
from a transport peak in the spectral functions. From that quantity we roughly estimated the heavy quark diffusion constant $D$ as
$2\pi DT \simeq$ 0.6--4 and 0.5--2 on our coarser and finner lattices, respectively, with quark mass in a range 1--2 GeV in the charm sector.
In the bottom sector, however, we couldn't estimate $D$ within 4--5 GeV quark mass.

Directly investigating the charmonium and bottomonium spectral functions, taking the continuum limit and developing a way to estimate the heavy
quark diffusion constant with much smaller systematic uncertainties are our future work.

\acknowledgments{
The numerical calculations have been performed on the Bielefeld GPU cluster and the OCuLUS Cluster
at The Paderborn Center for Parallel Computing in Germany.}

\end{document}